# A Comparison of Value-Added Models for School Accountability

George Leckie and Lucy Prior

Centre for Multilevel Modelling and School of Education, University of Bristol

**Corresponding author:**

Professor George Leckie

Centre for Multilevel Modelling

School of Education

University of Bristol

35 Berkeley Square

Bristol

BS8 1JA

United Kingdom

g.leckie@bristol.ac.uk

**Funding:**

This work was supported by the Economic and Social Research Council under Grant number ES/R010285/1.

**Acknowledgements:**

This work contains statistical data from Office for National Statistics (ONS), UK. The use of the ONS statistical data in this work does not imply the endorsement of the ONS in relation to the interpretation or analysis of the statistical data.

# A Comparison of Value-Added Models for School Accountability

## Abstract

School accountability systems increasingly hold schools to account for their performances using value-added models purporting to measure the effects of schools on student learning. The most common approach is to fit a linear regression of student current achievement on student prior achievement, where the school effects are the school means of the predicted residuals. In the literature, further adjustments are usually made for student sociodemographics and sometimes school composition and 'non-malleable' characteristics. However, accountability systems typically make fewer adjustments: for transparency to end users, because data is unavailable or of insufficient quality, or for ideological reasons. There is therefore considerable interest in understanding the extent to which simpler models give similar school effects to more theoretically justified but complex models. We explore these issues via a case study and empirical analysis of England's 'Progress 8' secondary school accountability system.

# Introduction

## School Systems Around the World

School systems around the world increasingly monitor and hold schools to account for their performances using school value-added models (VAMs) and student standardised tests (Organisation for Economic Co-operation and Development, OECD, 2008). These models purport to measure the effects schools have on student learning (the value that they add). High-profile examples include state and national accountability systems in Australia (Australian Curriculum, Assessment and Reporting Authority, ACARA, 2021), Chile (Muñoz-Chereau et al., 2020), England (Department for Education, 2020; Leckie & Goldstein, 2017), and many US states (Amrein-Beardsley, 2014; Koretz, 2017). Typically, the school effects are used to rank schools or place them in performance bands. This data is then returned to schools for self-evaluation and to promote school improvement. This information is also often used to inform the timing and focus of school inspections and high-stakes reward and sanction decisions for schools, including change of senior leadership and sometimes school takeovers or closures. Notably, in England the school effects are also published in 'league tables' to hold schools up to public scrutiny and to support parental school choice.

## School Value-Added Models

School value added models were first developed by the school effectiveness literature for the purpose of studying school policies and practices that might explain variation in school effectiveness (Mortimore et al., 1988; Reynolds et al., 2014; Rutter et al., 1979; Sammons et al., 1997; Teddlie & Reynolds, 2000; Townsend, 2007). They have also been widely explored by the educational statistics literature, especially from the perspective of holding schools to account for student learning (American Statistical Association, 2014; Braun & Wainer, 2006;

McCaffrey et al., 2004; Wainer, 2004). While various VAM approaches are used across different accountability systems (Castellano & Ho, 2013), the most common approach is to fit a linear regression of student current achievement on student prior achievement, where the school effects are the school means of the predicted residuals (Goldstein, 1997). Statistically adjusting for different combinations of student prior achievement and other student and school characteristics deemed beyond the control of schools (via entering covariates into the regression) leads to different predicted school effects. This has led to five main VAMs for school accountability, each with their own advantages and disadvantages.

*Raw Model*

The very simplest model is one which makes no statistical adjustments at all. The resulting school effects are effectively simple averages of student current achievement at the end of the phase of schooling or other period of interest. These models have been referred to as both 'Raw' and 'Type 0' models. We shall refer to them as 'Raw' models. Raw models should not be used to measure school effectiveness as by ignoring initial student achievement they fail to separate the value that schools add to student learning from pre-existing differences in learning across schools at the start of the phase of schooling. For this reason, the Raw model would also not usually be referred to as a VAM.

*VA Model*

The simplest true VAM is a linear regression of student current achievement at the end of the phase of schooling regressed on student prior achievement at the start of the phase (Aitkin & Longford, 1986; Goldstein, 1997; Raudenbush & Willms, 1995). The estimated regression coefficients describe the overall relationship between student current and prior achievement. Schools are then held accountable for how their students' scores systematically deviate from this. The school effects purport to measure the mean increase in student learning in each school over the phase (value-added period) relative to the average school. Schools with positive effects and whose 95% confidence intervals lie above zero are described as preforming "significantly above average", while schools with negative effects and whose 95% confidence intervals lie below zero are described as performing "significantly below average". These models have been referred to as both 'value-added' (VA) and 'Type AA' models. We shall refer to them as VA models.

*CVA-A Model*

The VA model is often extended to include student sociodemographic characteristics since these also vary across schools at intake, are similarly argued beyond the control of the school, and predict current achievement over and above prior achievement (Ballou et al., 2004; Leckie and Goldstein, 2019; Raudenbush & Willms, 1995). The resulting models have been referred to as both 'contextualised value-added' (CVA) and 'Type A' models. We shall refer to them as 'CVA-A' models. When prior achievement is omitted from these models, they are instead referred to as 'contextualised achievement' (CA) models (Lenkeit, 2013; OECD, 2008). A concern with both CA and CVA models is that they adjust away the national underperformance of student groups which are struggling, preventing schools from being directly challenged as to these disparities (Leckie & Goldstein, 2017).

*CVA-B Model*

The CVA-A model is sometimes further extended to adjust for school mean prior achievement and occasionally school means of one or more student sociodemographic covariates. Conceptually, these effects are added to capture school 'compositional' effects, specifically positive peer effects associated with being educated among higher prior achieving or otherwise more advantaged peers (Raudenbush & Willms, 1995; Timmermans & Thomas, 2015). These CVA models are sometimes referred to as 'Type B' models. We shall refer to them as 'CVA-B' models. Here too, support in the literature is not without caveats. Only some studies find significant effects, and these are often small (Thomas, 2001). Positive peer effects may be biased upwards if higher prior achieving students select into more effective schools (or more effective teachers choose to work in schools with higher prior achieving students). This in turn overadjusts the school effects making schools appear more homogenous and therefore a less important explanation for variation in student learning than they really are (Castellano et al., 2014; Prior et al., 2021).

*CVA-X Model*

Lastly, 'non-malleable' school characteristics are sometimes entered into VAMs with these also argued outside the control of the school, at least in the short run (Keeves et al., 2005). Such characteristics might include school size, type, and location. These CVA models have been referred to as 'Type X' models. We shall refer to them as 'CVA-X' models. Similar concerns arise as before around estimating potentially biased effects of these characteristics due to their likely correlations with the school effects. More fundamentally, for the purpose of identifying schools which are struggling to boost student learning, such adjustments have been argued as overadjustments. For example, adjusting for regional differences in student performance will lead the resulting school effects to be deviations from regional averages rather than the overall average, potentially masking which schools are underperforming nationally (Timmermans et al., 2011).

**School Accountability Systems**

School accountability systems wishing to hold schools accountable for student learning are faced with a choice as to which VAM to choose and more specifically which student and school characteristics to statistical adjust for. There are multiple reasons why they go on to make different choices.

First, while the school effectiveness literature is unanimous on the need to adjust for student prior achievement, choices still need to be made as how to measure prior achievement. Decisions must be made around the timing of this measurement as this defines the start and length of the value-added period and choices must be made around whether prior achievement is measured by a single academic subject or some aggregate measure across multiple subjects or whether some more general measure of ability or aptitude is used (Marks, 2017, 2021). Accountability systems will often be pragmatic, making use of whatever prior achievement tests are already in place (including teacher assessments if no standardised test).

Second, there is no consensus regarding the need to adjust for student sociodemographics, and less unanimity still regarding the need to adjust for school compositional variables, let alone non-malleable school characteristics. Thus, even if a school system wants to follow best practice according to the school effectiveness literature, there is ambiguity as to what this is.

Third, the users of VAMs in accountability systems include policy makers, schools, and the public, not just academic researchers, and there will be a natural desire for the chosen model to be simple enough to be easily understood and widely accepted. An obvious route to this is to omit potential adjustments.

Fourth, even if an accountability system wishes to use a VA model, they might not have the data to support this. For example, they may only run standardised tests at the end of a phase of education, not at the beginning, preventing adjustment for student prior achievement. Even if standardised tests at the beginning of the phase of education are introduced, it will still take several years before the first cohort of test takers reaches the end of the phase. Alternatively, a school system may only collect limited or no data on student sociodemographics. Where student prior achievement or sociodemographic measures are missing it follows that school compositional variables cannot be derived and so will also be omitted from such models.

Fifth, where a system is successfully using a VA or CVA model, events may arise which lead either the start or end-of-phase tests to be cancelled, for example due to boycotts, strikes, malpractice (leaked tests), or even a pandemic (COVID-19). Where end-of-phase tests are cancelled, there seems no obvious solution. But where start-of-phase tests are cancelled, an important question is then whether the school system could switch for that cohort only from a VA or CVA model to a CA model. Alternatively, systems might explore whether teacher assessments or an even earlier measurement of prior achievement could be substituted for the missing start-of-phase measure.

Sixth, under situations where information on prior achievement is available, there is the question of whether there is any benefit to additionally adjusting for an earlier achievement measure. For example, if students who progress rapidly between their early and prior achievement tests continue progressing rapidly between their prior and current achievement tests, then adjusting for early achievement would be expected to impact the school effects, adjusting down the school effects for schools with disproportionate numbers of such students. There is however little research on this.

Seventh, where a CVA model is desired and all the desired prior achievement and sociodemographic variables are available, there may still be data quality issues leading school systems to drop one or more variables from their models. For example, some potential prior achievement measures may be viewed as less reliable (due to measurement error) than others and so excluded. Additionally, where student special education needs are measured by the school as opposed to an official external organisation, there may be concerns over whether this is measured consistently across schools, leading the system to decide it is preferable to omit this adjustment altogether.

Eighth, ideological reasons may lead a system to omit certain adjustments. For example, some systems have argued that adjusting for student sociodemographics implicitly 'entrenches' lower expectations for educationally disadvantage students and this is 'morally wrong' (Department for Education, 2010, p. 68).

While the focus of the current study is on choice of VAM in the context of school accountability, and more specifically choices around which statistical adjustments should be made, there are of course many other issues which arise when specifying, estimating, and interpreting VAMs which can also impact the school effects. Prior et al. (2021) provide a non-technical review of the literature on these issues and, as we do in this article, discuss them in the context of the Progress 8 school accountability system in England. In brief, these issues include choice of student current achievement measure, including potential separate reporting by academic subject and inclusion of non-academic outcomes. They include decisions around which schools and students are excluded from VAMs, both intentionally due to their characteristics or planned random sampling, and unintentionally due to data linkage failing, student mobility, or missing data. Further concerns are around the presentation of school effects to users including challenges of communicating effect sizes and statistical uncertainty to users. Choice of statistical model is important including conventional versus multilevel linear regression as well as issues around allowing school effects to vary by student characteristics and reporting 'shrunken' school effects. Finally, issues related to the volatility of school effects over time should also be considered, including scope for reporting multi-year averages (via pooling students across consecutive cohorts).

**Previous Comparison Studies**

Given different VAMs make different adjustments, there is substantial interest in understanding the extent to which this might matter in practice for school accountability systems and whether simpler models might reasonably be substituted for more theoretically justified but complex models. Put simply, do the school effects change meaningfully with different models? Would different schools be judged as effective and ineffective? We focus here on five previous VAM comparison studies which have explored these questions via studying the impact of model choice on the magnitude of the school effects and their correlations across models (Leckie & Goldstein, 2019; Marks, 2017; Marks, 2021; Thomas & Mortimore, 1996; Timmermans et al., 2011). We also include two further comparison studies which, while not reporting the impact on the school effects, compare different VAMs in terms of their choice of adjustments (Muñoz-Chereau & Thomas, 2016; Thomas, 2001).

Collectively these seven studies compare seven main VAMs (Table 1), but individually each study only compares a subset of these, and none consider models which include an early measure of achievement either as a proxy for a missing prior achievement measure or as a complement to an observed prior achievement measure. Importantly, the studies make different recommendations as to best practice for accountability systems. These differences reflect not just the different conceptual arguments made across these studies, but differing empirical findings with the later relating in part to the different education systems, achievement tests, and points in time in which they were conducted as well as data availability and quality. Marks (2017) in their study of Australian schools recommends VA models over CA models arguing that the latter are a poor proxy for the former. In a follow-on study, Marks (2021) recommends accountability systems also focus on VA models over more complex CVA-A models, arguing that the latter make little difference in practice and that their additional adjustments can be susceptible to interpretational and data quality concerns. In contrast, Thomas and Mortimore (1996) in their study of schools in Lancashire, England, recommend CVA-A models over VA models noting the improvement in model fit associated with adjusting for student characteristics, but not school context measures. Leckie and Goldstein (2019), who study the Progress 8 school accountability system in England, also recommend CVA-A models over VA models, arguing that adjusting for student sociodemographics is necessary to avoid punishing and rewarding schools for simply serving socioeconomically disadvantaged or advantaged intakes. Thomas (2001), Timmermans et al. (2011) and Muñoz-Chereau & Thomas (2016) in their studies of English, Dutch and Chilean schools go further, recommending CVA-B models which additionally adjust for school compositional variables, arguing that these models prove best fitting in their studies and that schools should not be punished and rewarded for peer effects arising from the nature of their student intakes rather than the effectiveness of their teaching.

TABLE 1 SHOULD BE INSERTED AROUND HERE

**What we do in this Article**

In this article, we revisit the conceptual arguments for using different VAMs for holding schools to account and illustrate the consequences on the school effects. This work is important given the conflicting recommendations made by previous comparison studies and the high stakes attached to high and low school effects in many accountability systems. Our ultimate aims are therefore to inform and improve the design of such systems and to contribute to the school effectiveness literature more generally.

We extend previous VAM comparisons in four ways. First, we compare a wider range of VAMs than considered by any single previous study. Second, we explore the impact of omitting prior achievement from these models and the potential role of an early achievement measure both as a substitute for missing prior achievement and as a complement when prior achievement is observed. Third, we situate our study in the context of the current Progress 8 secondary school accountability system in England and use this as a case study to illustrate the arguments for and against different adjustments as well as to provide new empirical evidence. Fourth, the data are very large and detailed, a census of all students in all state-funded schools in England: 502,899 English secondary school students in 3,196 schools who completed schooling in 2019. This allows a deeper exploration of the impact of different adjustment variables on the school effects for both individual schools and for groups of schools with different characteristics.

## School Accountability in English Secondary Education

### A Long History

England has a long history of using VAMs to hold state-funded secondary schools to account for student learning stretching back thirty years. In contrast, private-funded secondary schools, which educate some 7% of all students, have always been excluded from this form of accountability. Saunders (1999) and Schagen et al. (2006) discuss the origins and early years of the VAM movement in England, while Leckie & Goldstein (2017) review and statistically critique the series of official VAM measures introduced from 2002 through 2016. The school effects from these models inform the timing and focus of school inspections and are additionally published in national tables (https://www.compare-school-performance.service.gov.uk/) to hold schools publicly accountable and to assist parents choosing schools (Burgess et al., 2019).

### Progress 8 School Effects Model

The current headline measure of school performance, "Progress 8", was introduced in 2016 (Leckie & Goldstein, 2017). The underlying model is a VA model. Specifically, a conventional linear regression of student Attainment 8 current achievement scores in national end-of-secondary school General Certificate of Secondary Education (GCSE) examinations (age 16, academic year 11, a total exam score across eight largely prescribed subjects) on a flexible function of student Key Stage 2 (KS2) prior achievement scores in national end-of-primary school tests (age 11, academic year 6, an average score across English and mathematics discretised into 34 groups and then entered as series of dummy variables). The school effects, referred to as school Progress 8 scores, are calculated as the school mean predicted residuals from this model. Scores are published with 95% confidence intervals to communicate their statistical uncertainty.

**Previous Value-Added Models**

Prior to 2016, the headline measure of school performance was a simple average measure of GCSE achievement scores (specifically a threshold measure: the percentage of students achieving five or more "good" GCSE exam grades where good is a grade in the range A*-C). These school effects are essentially the ones that arise from a Raw model. From 2004, these raw scores were complemented with various value-added scores. Most notably, from 2006 to 2010, a CVA-B model referred to simply as "CVA" was used which closely followed recommendations from the school effectiveness literature (Ray et al., 2009). The model adjusted for student prior achievement (KS2 score entered as a cubic polynomial), student sociodemographics (age, gender, ethnicity, language, special educational needs (SEN), free school meal (FSM), deprivation) including various interactions (ethnicity and FSM), and school compositional variables (school mean and SD of student prior achievement). The model was also a multilevel linear regression model and so included a school random intercept effect to directly measure the school effects. However, in 2011 this model was replaced with a simpler value-added model which only adjusted for student prior achievement. This change reflected a concern that the student sociodemographic adjustments were entrenching low aspirations for disadvantaged students (Department for Education, 2010). In 2016, with the introduction of Progress 8, the underlying model was additionally changed to a conventional linear regression model as the latter was viewed as simpler to explain to policy makers and schools (Burgess & Thomson, 2013).

**Implications of COVID-19 Exam Cancellations**

The COVID-19 pandemic led the 2020 and 2021 GCSE examinations to be cancelled and so Progress 8 was not published for these two years. KS2 tests were also cancelled meaning Progress 8 cannot be published in 2025 or 2026 when the effected students reach the end of their secondary schooling. This raises an important question: Can a different VAM, not reliant on student prior achievement, produce similar school effects and therefore be used for these two years. In particular, can an earlier measure of student achievement, student Key Stage 1 (KS1) score in teacher assessments midway through primary schooling (age 7, academic year 2, an average score across English and mathematics), be used as a proxy for missing KS2 prior achievement? What if student sociodemographics are added to the model in place of prior achievement? Finally, it should be noted that this is not the first time that the KS2 prior achievement tests have been disrupted. In 2010, a quarter of primary schools boycotted the KS2 tests (BBC News, 2010) preventing school value-added measures from been reliably calculated in 2015.

## Methods

**Data**

We analyse the official data that underlie England's Progress 8 secondary school accountability system. These data are drawn from the National Pupil Database (Department for Education, 2021).

**Sample**

The sample is the same as that used to calculate the 2019 Progress 8 measure and consists of 502,899 students in 3,196 schools. School size ranges from 6 to 544 students with a mean of 157.

**Measures**

*Student current achievement* is measured by student GCSE Attainment 8 score in 2019. We standardize Attainment 8 to have mean 0 and SD 1 to make the results accessible to an international audience.

*Student prior achievement* is measured by their KS2 score in 2014 and, following the Progress 8 methodology, is discretised into 34 achievement bands to flexibly capture the non-linear national relationship between current and prior achievement (monotonically increasing and convex).

*Student early achievement* is measured by their KS1 score in 2010 and, to be consistent with the Progress 8 treatment of the KS2 score, we also discretise this variable, choosing 20 achievement bands (KS1 scores are less finely granulated than KS2 scores). Students with missing KS1 scores (4.24%) were assigned to an additionally missing category (no other variables have missing values).

*Student sociodemographics characteristics* are age, gender, ethnicity, language (whether they speak English as an additional language), SEN, FSM, and deprivation decile associated with student postcode (zip code). These are the same characteristics that were used in the 2006-2010 CVA model.

*School compositional variables* are the school mean prior achievement band and early achievement band measured by ventiles (20 equal sized groups) of school mean student KS2 achievement and KS1 achievement respectively.

*Non-malleable school characteristics* are region, type, admissions policy, age range, gender, religious denomination, and deprivation decile associated with the school postcode. We note that for some school characteristics their classification as non-malleable is clearer than for others, a known tension when specifying CVA-X models (Keeves et al., 2005). For example, schools can apply to change their type, though this might take several years. Similarly, for some school characteristics, their classification as non-malleable versus compositional is debateable. For example, school deprivation might reasonably by conceptualised as a school compositional variable in which case we would derive it as the school mean of the student deprivation scores.

Descriptive statistics for all measures available on request.

**Method of Analysis**

We fit all models as conventional rather than multilevel linear regression models as this is the methodology used for Progress 8 and an important aspect of our work is to situate it within the context of a real accountability system setting replicating the choices that they make. Regression coefficients and school effects from multilevel linear regression versions of these models give similar results (Prior et al., 2021). However, this will not necessarily be the case in other studies. In particular, school effects models fitted in studies characterised by smaller schools and stronger school effects will be more likely to show divergent results for multilevel and conventional linear regression and, in those settings, the multilevel results would typically be viewed more favourably. We specify cluster robust standard errors to account for school-level residual clustering. The school effects are calculated as the school means of the student predicted residuals. We also calculate their 95% confidence intervals.

**School Value-Added Models**

Student current achievement is the outcome in all models. We start by fitting the most common models introduced above: Raw, VA, CVA-A, CVA-B and CVA-X (see Introduction and Table 1). We enter covariates in line with Progress 8 and previous VAM analyses of these data and for simplicity we do not include interaction terms. We then refit this sequence of models three times, altering the nature of the adjustment for student prior achievement, the most important of all the adjustment variables. In the first re-fit, we omit prior achievement altogether and so fit contextualised versions of all models. This mimics the scenario where accountability systems are yet to reach the first student cohort where prior achievement can be matched to current achievement. In the second re-fit, we substitute early achievement for prior achievement, mimicking the scenario where the usual prior achievement measure proves unusable for a given cohort. In the third ref-fit, we enter prior and early achievement simultaneously to explore whether there are benefits to adjusting for early achievement over and above just prior achievement.

**Comparing Models**

We first compare models in terms of their estimated regression coefficients and adjusted R-squared statistics. We then compare models in terms of the practical significance (effect size) and statistical significance of their school effects. Last, we compare models in terms of how associated their school effects are via Pearson and Spearman correlations and scatterplots.

<div align="center"><b>Results</b></div>

**Comparing School Value-Added Models: Raw, VA, CVA-A, CVA-B, CVA-X**

*Estimated Regression Coefficients and Adjusted R-squared*

Our primary interest is in comparing the school effects across the Raw, VA, CVA-A, CVA-B, and CVA-X models. We therefore only briefly summarise the regression coefficients and adjusted R-squared statistics for each model here. Model summary results are given in Table 2. Full model results for these and subsequent models are available on request.

TABLE 2 SHOULD BE INSERTED AROUND HERE

The Raw model regresses student current achievement on only an intercept and so has an adjusted R-squared of .000.

The VA model adjusts for student prior achievement (dummy variables entered for 34 groups of KS2 student achievement). The regression coefficients show the expected strong positive convex relationship between current and prior achievement (the model is equivalent to the Progress 8 model). The adjusted R-squared is .542 and so student prior achievement explains over half the variation in student current achievement. Subsequent models also show that student prior achievement is by far the most important predictor of current achievement. Therefore, our findings with a recent, large national sample are in agreement with those long established in the literature (e.g., Thomas and Mortimore, 1996).

The CVA-A model additionally adjusts for the seven student sociodemographic characteristics (age, gender, ethnicity, language, SEN, FSM, and deprivation). The adjusted R-squared increases from .542 to .610 and so student sociodemographics meaningfully predicts student current achievement even after adjusting for student prior achievement. Younger students, girls, nearly all ethnic minority groups, students who speak English as an additional language, students without SEN, students not on FSM, and students living in less deprived neighbourhoods are all predicted higher current achievement than otherwise equal students. This pattern of results is also long-standing, agreeing with previous studies of student achievement in England that have explored similar characteristics (e.g., Leckie and Goldstein, 2009, 2019; Ray et al., 2009; Thomas, 2001; Thomas and Mortimore, 1996).

The CVA-B model additionally adjusts for school mean prior achievement (ventiles of school mean student KS2 achievement band) and shows a positive relationship. However, the adjusted R-squared only increases from .610 to .614 suggesting the positive peer effects associated with being educated among higher prior achieving peers are of little practical importance. Furthermore, these positive effects do not increase monotonically across the 20 ventiles. Marks (2021) and Muñoz-Chereau & Thomas (2016) also report positive effects of school mean prior achievement in their study. However, Thomas (2001) reports only a small improvement in goodness of fit from including school mean prior achievement in their CVA-B model. Thomas and Mortimore (1996) report that none of their school context measures are significant, though this may reflect the low number of schools and therefore power in their study, just 79 schools versus over 3000 schools in the current study.

Finally, the CVA-X model additionally adjusts for the seven non-malleable school characteristics (region, type, admissions policy, age range, gender, religious denomination, and deprivation decile). All adjustments, except deprivation, are statistically significant. The non-significant effect of school deprivation is perhaps not so surprising given that the model has already adjusted for student FSM and deprivation. The regression coefficients suggest that students in London schools, those in converter academies, those in grammar schools, students in single-sex schools, and those in religious schools, are all predicted higher current achievement than otherwise equal students. However, the adjusted R-squared only increases from .614 to .618 suggesting that, as with school mean prior achievement, having adjusted for student prior achievement and sociodemographics, non-malleable school characteristics are not very important for further explaining student current achievement. In their study, Timmermans et al. (2011) report an even more extreme finding that none of their CVA-X school characteristics prove statistically significant, though this might also reflect low power due in this case to just 60 schools.

*School Effects: Effect Size and Statistical Significance*

Recall, we standardized the Attainment 8 measure of student current achievement to have mean 0 and variance 1. The Raw model makes no adjustments and so the variance of the student residuals is also 1. The variance of the school mean residuals or school effects is 0.226 and so represents 22.6% of the variance in student current achievement. Thus, there is substantially less variation in students' current achievement scores between schools than there is within schools, a point that is sometimes lost in public discussions of school effects: schools are important, but only tell part of the overall story. It is of interest to compare this and subsequent results to those from previous comparison studies (see Table 1). To facilitate this, we present the school effect variances and correlations between school value-added models reported by the seven earlier studies in a Supplemental File. The variation shown in these statistics will reflect the education system, nature of achievement tests, time period, data considerations, and many other factors specific to each study. Marks (2021), Muñoz-Chereau & Thomas (2016), and Timmermans et al. (2011) report corresponding statistics of 30%, 37% and 7% in their studies suggesting the relative importance of schools as a source of variation in student achievement varies greatly across countries.

The variance and therefore magnitude of the school effects reduces substantially across the five models (0.226, 0.069, 0.043, 0.039, 0.032) as more adjustments are made for factors argued beyond the control of the school. This suggests that simpler models overstate the impact schools have on student learning. Moving from the Raw to the VA model sees by far the biggest decrease in variance from 0.226 to 0.069 suggesting that, once we account for school intake differences in student prior achievement, schools only account for a much lower 6.9% of the variation in student current achievement. Additionally adjusting for sociodemographics lowers the variance further to 0.043 or just 4.3% of the variation in student current achievement. Further, but smaller reductions in the variance are then seen as we move to the CVA-B (0.039) and CVA-X models (0.032). This patterning of results whereby the reductions in the school variance diminish as we move across the five models with the biggest drop associated with student prior achievement is in agreement with previous comparison studies (Marks 2021; Muñoz-Chereau & Thomas, 2016; Thomas & Mortimore, 1996) (see Supplemental File).

To better understand what impact choice of VAM has on the magnitude of the individual school effects, we define schools as having 'no or very small', 'small', 'moderate', or 'large' school effects when their school effects take absolute values in the ranges 0 to 0.1, 0.1 to 0.2, 0.2 to 0.45, or greater than 0.45, respectively, where these values are measured in SD units of student current achievement. School effects with 95% confidence intervals overlapping 0 are automatically placed in the no or very small category. These definitions approximately map those used to assess school-based randomised control trials and other interventions in England (Education Endowment Foundation, 2021). Figure 1 then presents a stacked bar chart of the percentage of schools whose effects fall into each category separately for each model. The percentage of schools with moderate or large effects decreases across the five models (62%, 40%, 28%, 25%, 24%). As expected, the biggest decrease is seen when moving from the Raw to the VA model (62% to 40%), consistent again with student prior achievement being the single most important adjustment. However, the decrease associated with moving from VA to CVA-A is still substantial (40% to 28%) demonstrating that adjusting for student sociodemographics has a real impact on how many individual schools would be viewed as having moderate or large effects.

FIGURE 1 SHOULD BE INSERTED AROUND HERE

*School Effects: Correlations and Scatterplots*

We now turn our attention to the correlations and scatterplots between the five sets of school effects to understand the impact of each set of covariate adjustments on the relative performance of schools. Pearson and Spearman correlations are similar and so we restrict our discussion to the former (Table 3). All correlations are positive and high (range .41 to .96) as to be expected given all models purport to measure school effects on student learning, albeit in different ways. They are also highest for adjacent models, VA vs. Raw, CVA-A vs. VA, CVA-B vs. CVA-A, and CVA-X vs. CVA-B, since each adjacent pair of models differs only by a single set of adjustments (.81, .90, .96, .91). The lowest correlation is for the two models which differ most in their adjustments, Raw and CVA-X (.41). This patterning of results and approximate magnitudes is in line with those presented in previous comparison studies (Leckie and Goldstein, 2019; Marks, 2017; Marks 2021; Thomas & Mortimore, 1996; Timmermans et al., 2011) (see Supplemental File).

TABLE 3 SHOULD BE INSERTED AROUND HERE

Henceforth, we focus on comparing adjacent models since these represent the most likely comparisons of interest to accountability systems. To support our discussion, Figure 2 presents scatterplots of the school effects (left column) and ranks (right column) for each pair of adjacent models (rows). These plots show that many individual schools are dramatically affected by each set of adjustments despite the high correlations. The plots also reveal subgroups of schools that are affected quite differently from the majority of schools. With the exception of Leckie and Goldstein (2019), the seven previous comparisons studies did not present such plots, but these results highlight the importance of doing so.

FIGURE 1 SHOULD BE INSERTED AROUND HERE

The top row compares VA and Raw school effects. The correlation is .81. Thus, schools with the highest mean current achievement are in general still the schools which appear most effective once student prior achievement is adjusted for, but this relationship is by no means guaranteed. Many schools' scores and ranks change considerably across the two models. Importantly, the plot reveals grammar schools (162 schools or 5% nationally) appear far less impressive in terms of their VA school effects than they do in terms of their Raw school effects. Grammar schools are unusual in that they set entrance examinations and are therefore characterised by very high prior achieving students. In contrast to these results, Timmermans et al. (2011) and Marks (2021) report somewhat higher correlations of .95 and .89 (see Supplemental File). While it is tempting to draw conclusions from this relating to differences between the school systems, we note that Leckie and Goldstein (2019) in their analysis of the Progress 8 accountability system reported a lower correlation of .75 between VA and Raw school effects for the student cohort just three years earlier than those analysed in the current study.

The second row compares CVA-A and VA school effects. The correlation is now higher at 0.90 suggesting that while additionally adjusting for student sociodemographics changes the measured relative performance of schools, this change proves less dramatic than that associated with adjustment for prior achievement. Furthermore, additionally adjusting for sociodemographics does not appear to consistently reduce the relative effectiveness of grammar schools further; some grammar schools appear more effective, other less effective relative to other schools. This may seem surprising given the known socioeconomically advantaged nature of grammar school intakes (Leckie & Goldstein, 2019). The explanation is that grammar school students are disproportionately white British, but nationally white British students are associated with underperformance. Thus, while adjusting for student socioeconomic status (FSM status and deprivation decile) leads grammar schools to appear less effective, this is countered by an increase in effectiveness resulting from adjusting for ethnicity. Marks (2021), Leckie and Goldstein (2019), Timmermans et al. (2011), and Thomas and Mortimore (1996) report correlations of .72, .91, .82, .94 in their comparisons of CVA and VA effects (see Supplemental File). Thus, the impact of adjusting for student sociodemographics appears somewhat greater in the Dutch and Australian contexts than in the English context, though we again caution against overinterpreting such differences.

      The third row compares CVA-B and CVA-A school effects. The correlation is now very high at 0.96 suggesting that additionally adjusting for positive peer effects associated with higher school mean prior achievement makes comparatively little difference to the relative performance of the majority of schools. While the plots largely confirm this, grammar schools are again affected quite differently from schools in general due to their unusually high prior achieving students. Specifically, grammar schools make up 97% of the 159 schools in the highest ventile of school mean prior achievement. This illustrates an important concern we have raised with attempts to adjust for peer effects. If grammar schools are truly more effective than standard schools, then the selection of higher prior achieving students into grammar schools will lead the positive peer effects estimated by the CVA-B model to be biased upwards in which case the school effects will be biased towards zero for all schools, but especially so for grammar schools (Leckie & Goldstein, 2009). We note that the plot also reveals a second subgroup of schools which are also notably negatively affected by adjusting for school mean student prior achievement, though not as dramatically as for grammar schools. This second subgroup relates to schools in the second highest ventile of school mean prior achievement, but in contrast to the highest ventile they are not dominated by any single school type. Our finding that adjusting for peer effects matters relatively little agrees with Marks (2021) and Timmermans et al. (2011) who also both report high correlations of .96 and .90 between CVA-A and CVA-B models (see Supplemental File).

      The fourth row compares CVA-X to CVA-B school effects. The correlation is lower at 0.91 revealing a further change in the relative performance of schools induced when

adjustments are made for non-malleable school characteristics. Here the plots shows that while grammar schools do not appear to be systematically affected by this set of adjustments, four other school types are. Further Education Colleges (FECs), University Technical Colleges (UTCs), and Studio schools all appear relatively more effective after adjustment. These schools typically only teach students from age 14 (for the final two years of secondary schooling), taking students that have often struggled in education up to this point. The emphasis in these school types is more focused on vocational education in preparation for future careers rather than the academic subjects prescribed by Attainment 8 (Prior et al, 2021). These points suggest that it makes little sense forcing these schools into the same accountability measure as for all other schools which teach students from age 11 (all five years of secondary schooling) and focus on academic subjects. In contrast, Muslim schools (only 14 schools or 0.4% nationally) appear less effective after adjustment for non-malleable school characteristics, in this case religious denomination. The plots illustrate a concern we have raised with the CVA-X model which is that it effectively redefines the school effect for each school as the effect of that school relative only to other schools of the same type. National differences in school effectiveness across different school types such as the higher average performance of Muslim schools is lost, yet these differences are part of the variation of interest to school accountability systems attempting to identify struggling schools. Timmermans et al. (2011) reports a somewhat higher correlation of .96 between their CVA-B and CVA-X school effects (see Supplemental File).

**Omitting Prior Achievement, Substituting Early Achievement for Prior Achievement, and Adjusting simultaneously for Early and Prior Achievement**

We now refit the original five models (Raw, VA, CVA-A, CVA-B, CVA-X) a further three times, first omitting prior achievement, next substituting early achievement for missing prior achievement, then simultaneously adjusting for early and prior achievement. Rather than go through the results for these new models in the same level of detail as we did for the original models, we describe at a summary level how these new results differ from before.

Figure 3 plots the percentage of schools with moderate or large effects (statistically significant effects greater than or equal to 0.2 SD of student current achievement) in each model separately for each possible choice of adjusting for student prior and early achievement. Figure 4 plots the Pearson correlation for each model between the school effects resulting from each possible choice of adjusting for student prior and early achievement and the original VAM.

FIGURE 3 SHOULD BE INSERTED AROUND HERE

FIGURE 4 SHOULD BE INSERTED AROUND HERE

First consider the original sequence of five models which includes student prior achievement (solid line in Figure 3) and which shows the percentage of moderate or large school effects decreases across the five models (61.9%, 40.4%, 28.1%, 25.1%, 23.4%). Relative to these models, omitting student prior achievement from each model (dashed line in Figure 3) dramatically increases the percentages of schools classified as having moderate or large effects (61.9%, 61.9%, 47.1%, 47.1%, 36.7%). The correlation between the resulting school effects and those of the original models (dashed line in Figure 4) are also substantially reduced (1.000, .815, .778, .584, .794). These results highlight how omitting student prior achievement leads to very different school effects whatever the preferred model. Different schools would be judged effective and ineffective.

Next, we consider substituting early achievement for missing prior achievement (dotted line in Figure 3). This considerably reduces the percentage of schools with moderate or large effects (61.9%, 48.8%, 39.0%, 33.8%, 28.3%) taking us approximately halfway back to the percentages seen for the original models. The correlations between the resulting school effects and those of the original models also increase (dotted line in Figure 4), but are still nowhere near high enough to suggest a similar ordering of school effects (1.000, .904, .883, .820, .862). These results suggest that early achievement does not make a good proxy for missing student prior achievement. This may in part relate to early achievement being teacher assessed and therefore less reliable than had it been derived from a test, a point we return to in the Discussion.

Last, we consider simultaneously adjusting for early and prior achievement (dash dot line in Figure 3). This reduces the percentage of schools with moderate or large effects only very slightly relative to the original models (61.9%, 38.7%, 27.9%, 25.0%, 23.0%). The correlations between the resulting school effects and those of the original models (dash dot line in Figure 4) are now also extremely similar (1.000, .994, .995, .988, .988). These results suggest that there is very little benefit to adjusting for early achievement in addition to prior achievement. The school effects are essentially the same.

Some specific further comparisons merit attention. The CVA-A model excluding prior achievement is an example of a CA model: a model which only adjusts for student sociodemographics. This model has a considerably higher percentage of moderate or large school effects than that for the original VA model including prior achievement (47.1 vs. 40.4) and the correlation between the CA and VA school effects is just .739 (not shown). This suggests that a CA model would also be a poor proxy for the VA model in situations where there is no prior achievement measure. This low correlation is similar to the low correlation of .77 reported by Thomas and Mortimore (1996) in their comparison of CA and VA school effects, but is somewhat lower the correlation of .85 reported by Marks (2017) (see Supplemental File). Furthermore, the school effects from the CA model are more strongly correlated with the Raw model school effects (.902; not shown). In contrast, the CVA-A model with early achievement substituted for missing prior achievement shows a similar percentage of moderate or large school effects to that for the original VA model including prior achievement (39.0 vs 40.4), however the correlation between the school effects is still just .811 (not shown). Thus, even simultaneously adjusting for student early achievement and sociodemographics proves a poor proxy to the original VA model which only adjusts for prior achievement.

## Discussion

### What are the Strengths and Weaknesses of Different Value-Added Models for School Accountability and Does Choice of Model Make a Difference in Practice?

Our starting point was to focus on the five VAMs most often discussed in the school effectiveness literature: the Raw, VA, CVA-A, CVA-B, and CVA-X models. In each model, the school effects are calculated as school mean residuals from a linear regression of student current achievement at the end of a phase of schooling. The models differ in the covariate adjustments they make for student and school characteristics measured at the start of the phase of schooling. The Raw model made no adjustments. The VA model adjusts for student prior achievement. The CVA-A model adds student sociodemographics. The CVA-B model adds school compositional variables (school means of the student characteristics). The CVA-X model adds non-malleable school characteristics.

Previous comparison studies have made conflicting recommendations regarding which model is most appropriate for school accountability purposes arguing in favour of VA models (Marks, 2017; Marks, 2021), CVA-A models (Leckie & Goldstein, 2019; Thomas and Mortimore, 1996), and CVA-B models (Muñoz-Chereau & Thomas, 2016; Timmermans et al., 2011; Thomas, 2001), respectively. As well as reflecting different conceptual arguments, these differing recommendations will also relate to varying empirical findings, resulting in part from differences in achievement tests, data availability and quality, as well as the different educational and temporal contexts in which they were conducted. Our own review and comparison of these models lead us to also reject the Raw and CVA-X models. Of the three remaining models we favour the VA and CVA-A models arguing that they each have their own strengths and weaknesses and more is to be learnt by presenting and contrasting their school effects side-by-side rather than choosing between them.

First, we agree with the wide held view that the Raw model is inadequate for school accountability purposes as it makes no adjustment for initial school differences in student achievement (Mortimore et al., 1988; Reynolds et al., 2014; Rutter et al., 1979; Sammons et al., 1997; Teddlie & Reynolds, 2000; Thomas & Mortimore, 1996; Townsend, 2007). Our results provided a compelling example of the importance of adjusting for prior achievement. In the VA model, the performance of grammar schools is no different from many non-grammar schools. In contrast, the Raw model misleadingly portrays grammar schools as unrivalled in their effectiveness as it fails to account for their students' unusually high prior achievement scores.

At the other extreme, the CVA-X model also seems inappropriate unless interest is limited to making comparisons within school groups defined by common non-malleable school characteristics (Timmermans et al., 2011). However, school accountability systems usually want to compare all schools to one another to identify schools that are struggling regardless of their school type. For instance, the VA and CVA-A models both highlight schools offering more vocational education as struggling (FECs, UTCs, and Studio schools). In contrast, the CVA-X model adjusts this national underperformance away, running the risk that these schools would be viewed as more effective than they truly are if their school effects are mistakenly compared to those of other school types.

On first inspection, the CVA-B model seems most conceptually appealing in attempting to account for potential peer effects as well as student achievement and sociodemographic characteristics that vary between schools (Muñoz-Chereau & Thomas, 2016; Timmermans et al., 2011; Thomas, 2001). However, if higher prior achieving or otherwise more educationally advantaged students select into more effective schools, these peer effects will be biased upwards leading the school effects to be biased towards zero (Castellano et al., 2014). In our study, moving from the CVA-A to CVA-B models resulted in grammar schools seeing their average performance dramatically pulled towards the national average and it seems plausible that this in part reflects the bias described here. However, there is no easy way to evaluate whether this bias is small or large. Peer and selection effects are confounded. Given this, our view is that it seems reasonable that school accountability systems might restrict their attention to simpler VA and CVA-A models even though these implicitly assume no peer effects at all.

In the CVA-A model, the adjustment for student prior achievement is typically far more important than the additional adjustments for student sociodemographics. When the resulting school effects change very little, the choice between VA and CVA-A becomes trivial, the simpler VA model is preferred. However, when the school effects change meaningfully, accountability systems face a harder choice. In our study, the CVA-A school effects were both smaller in magnitude than the VA school effects and weakly correlated enough to consequentially change the relative effectiveness of schools. While the desire to make like-with-like comparisons is, for many, an all persuasive argument in favour of adjusting for student sociodemographics, there are also important arguments against such adjustments and these should also be considered (Marks, 2021). First, the ability to make additional adjustments depends on data availability and quality. For example, in our CVA-A models we attempted to adjust for student socioeconomic status by adjusting for student

FSM, but the latter is a crude proxy for the former (Jerrim, 2020) and this likely has some impact on the resulting school effects. There may well become a point where data limitations are deemed so great that making no student sociodemographic adjustments is considered preferable to making inadequate adjustments. Second, the more adjustments that are made, the more complex the resulting model will be to end user: there becomes a point where conceptually justified complexity might pragmatically be sacrificed in the interests of accessibility. Third, it has been argued that CVA-A models, via adjusting for student sociodemographics, absolves individual schools for any national underperformance of disadvantaged students, whereas VA models hold individual schools wholly responsible for such patterns. Given the national underperformance of disadvantaged students reflects a complex interaction of both school and society influences, failing to simultaneously recognise both sets of influences seems unattractive. In sum, while there are strong arguments for CVA-A, there are also strong arguments against. Perhaps pragmatically, rather than picking one or other, accountability systems might aim to report both VA and CVA side-by-side, paying particular attention to explaining those schools which perform differently across the two measures (Thomas and Mortimore, 1996).

We acknowledged that, as with previous comparison studies, our recommendations reflect not just the conceptual arguments we make, but the context and empirical findings of our case study. Thus, the choices that accountability systems make may reasonably vary given the different contexts and data circumstances in which they operate. What is important is that systems carefully explain and explore alternatives to the choices that they make.

**What are the Consequences of Omitting Student Prior Achievement, Substituting Early Achievement for Prior Achievement, or Adjusting for Both Early and Prior Achievement?**

The literature is unanimous in that VAMs for school accountability should include student prior achievement and that this is the single most important adjustment to make. A key contribution of our study is to therefore consider what a school accountability system might do in the absence of such a measure, either because the system does not operate a prior achievement test, or if the usual prior achievement test had to be cancelled (e.g., due to boycotts, strikes, malpractice). In the context of Progress 8, COVID-19 led to the cancellation of the 2019 and 2020 prior achievement tests preventing Progress 8 from being calculated in 2025 and 2026 when these students take their end of secondary phase achievement tests.

We therefore studied the impact of omitting prior achievement from the VA, CVA-A, CVA-B and CVA-X models. The resulting school effects were fundamentally different both in magnitude (larger) and correlation with the original models (sufficiently low to suggest a meaningfully different rank ordering of schools). Comparing a CVA-A model without prior achievement (i.e., a CA model) to the VA model with prior achievement also suggested that student sociodemographics would make a poor proxy for missing prior achievement, a finding in agreement with Marks (2017). We also explored whether an early achievement measure could be used as a proxy for missing prior achievement. The resulting school effects, while now closer to those derived from VAMs with prior achievement, were still fundamentally different in their magnitude (still larger) and correlation (still not sufficiently close to 1). One explanation why the school effects were not closer is that the early achievement KS1 score was derived from teacher assessments and therefore potentially less reliable than had it been derived from standardised tests. In this case, the regression coefficient adjustment for early achievement would suffer from attenuation bias resulting in school effects of larger magnitude and differently correlated than they would otherwise be. Thus, in terms of calculating Progress 8 in 2025 and 2026, these results suggest that neither the seven available student sociodemographic characteristics nor students' KS1 scores would make good proxies for their missing Key Stage 2 scores, though making both these adjustments would certainly be better than falling back on the Raw model that makes no adjustments (Thomas & Mortimore, 1996). Last, we considered whether there is any benefit to simultaneously adjusting for early and prior achievement. Here the resulting school effects were very similar in both magnitude and correlation to those derived from the VAMs without early achievement suggesting that there is little benefit from including this extra measure.

**Tables**

**Table 1.**

Summary of school value-added models compared in previous comparison studies.

| | School effects model | | | | | | |
|---|---|---|---|---|---|---|---|
| | Raw | VA | CVA-A | CVA-B | CVA-X | VA-B | CA |
| Previous model comparison studies | | | | | | | |
| Marks (2017) | | **Models 2, 3, 4, 5** | | | | | Model 1 |
| Marks (2021) | Model 0 | **Models 1,2** | Model 3 | Model 4 | | Model 5 | |
| Muñoz-Chereau & Thomas (2016) | Raw | VA | | **CVA** | | | CA |
| Leckie & Goldstein (2019) | | P8 | **AP8** | | | | |
| Thomas (2001) | Model 1 | Models 3,4,5 | Model 6 | **Models 7, 8** | | | Model 2 |

| | | | | | | |
|---|---|---|---|---|---|---|
| Thomas & Mortimore (1996) | | Prior attainment | **Basic model, Refined model** | | | No prior attainment model |
| Timmermans et al. (2011) | Type 0 | Type AA | Type A | **Type B** | Type X | |
| Covariate adjustments | | | | | | |
| Student prior achievement | | × | × | × | × | × |
| Student sociodemographics | | | × | × | × | × |
| School mean prior achievement | | | | × | × | |
| Non-malleable school characteristics | | | | × | × | |

Notes.

Model names given in row 2 are those used in the current study. Model names given in rows 4-10 are those used in previous comparison studies. VA = value-added; CVA = contextual value-added; CA = contextualised attainment; P8 = Progress 8; AP8 = Adjusted Progress 8. Model names highlighted in bold are those recommended as the preferred model by each study. See each study for further details.

**Table 2.**

Model summary for the five main school value-added models: Raw, VA, CVA-A, CVA-B, CVA-X.

|  | School value-added models | | | | |
| --- | --- | --- | --- | --- | --- |
|  | Raw | VA | CVA-A | CVA-B | CVA-X |
| Statistical adjustments |  |  |  |  |  |
|   Student prior achievement |  | × | × | × | × |
|   Student sociodemographics |  |  | × | × | × |
|   School mean prior achievement |  |  |  | × | × |
|   Non-malleable school characteristics |  |  |  |  | × |
| Adjusted R-squared | 0.000 | 0.542 | 0.610 | 0.614 | 0.618 |
| SD of residuals | 1.000 | 0.677 | 0.624 | 0.621 | 0.618 |
| SD of school effects | 0.476 | 0.263 | 0.207 | 0.197 | 0.178 |
| Variance of residuals | 1.000 | 0.458 | 0.390 | 0.386 | 0.381 |
| Variance of school effects | 0.226 | 0.069 | 0.043 | 0.039 | 0.032 |
| % of residual variance due to schools | 22.6 | 15.1 | 11.0 | 10.1 | 8.3 |
| % of schools statistically significant | 70.9 | 66.9 | 63.1 | 58.9 | 57.7 |

Notes.

Sample size = 502,899 students in 3,196 schools. VA = value-added; CVA = contextual value-added.

**Table 3.**

Pearson (lower triangle) and Spearman (upper triangle) correlations between the school effects for the five main school value-added models: Raw, VA, CVA-A, CVA-B, CVA-X.

|       | Raw | VA  | CVA-A | CVA-B | CVA-X |
|-------|-----|-----|-------|-------|-------|
| Raw   | 1   | .84 | .70   | .52   | .46   |
| VA    | .81 | 1   | .89   | .81   | .72   |
| CVA-A | .70 | .90 | 1     | .94   | .87   |
| CVA-B | .49 | .84 | .96   | 1     | .91   |
| CVA-X | .41 | .71 | .87   | .91   | 1     |

Notes.

Sample size = 3,196 schools. VA = value-added; CVA = contextual value-added.

**Figure 1.**

Stacked bar chart of percentage of schools classified as having "no or very small", "small", "moderate", or "large" school effects, plotted separately by each school value-added model.

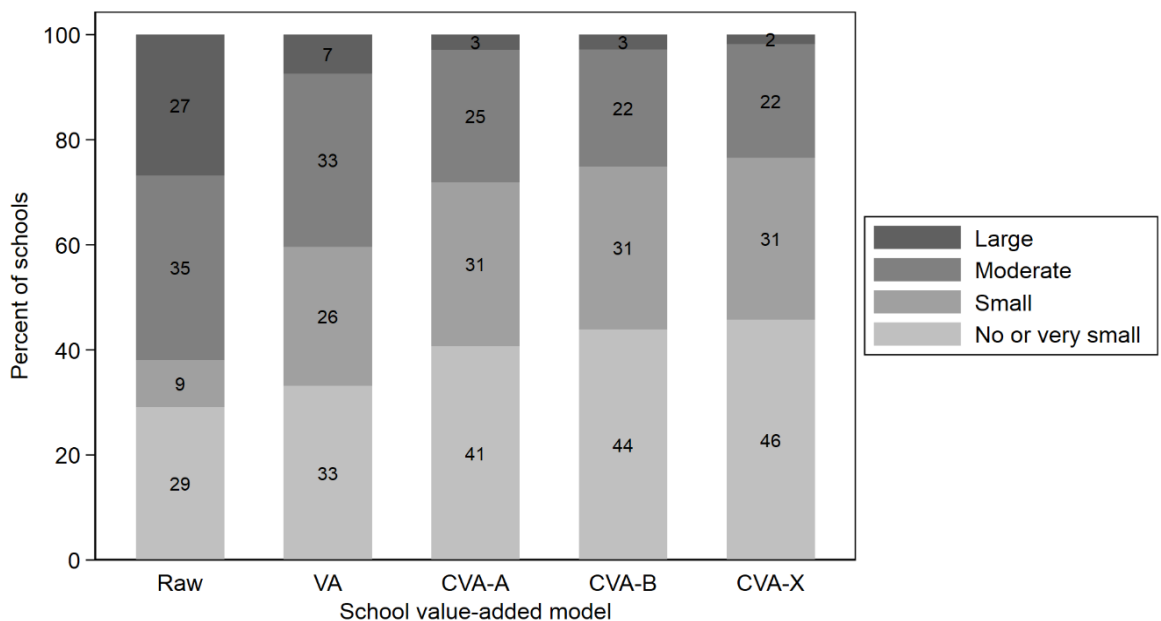

Notes.

Sample size = 3,196 schools. VA = value-added; CVA = contextual value-added. Schools are defined as having "no or very small", "small", "moderate", or "large" school effects when their school effects take absolute values in the ranges 0 to 0.1, 0.1 to 0.2, 0.2 to 0.45, or greater than 0.45, where these values are measured in SD units of student current achievement.

**Figure 2.**

Scatterplots of school effects (left column) and ranks (right column) for adjacent value-added models (rows) presented with Pearson and Spearman correlations: VA vs. Raw, CVA-A vs. VA, CVA-B vs CVA-A, CVA-X vs. CVA-B.

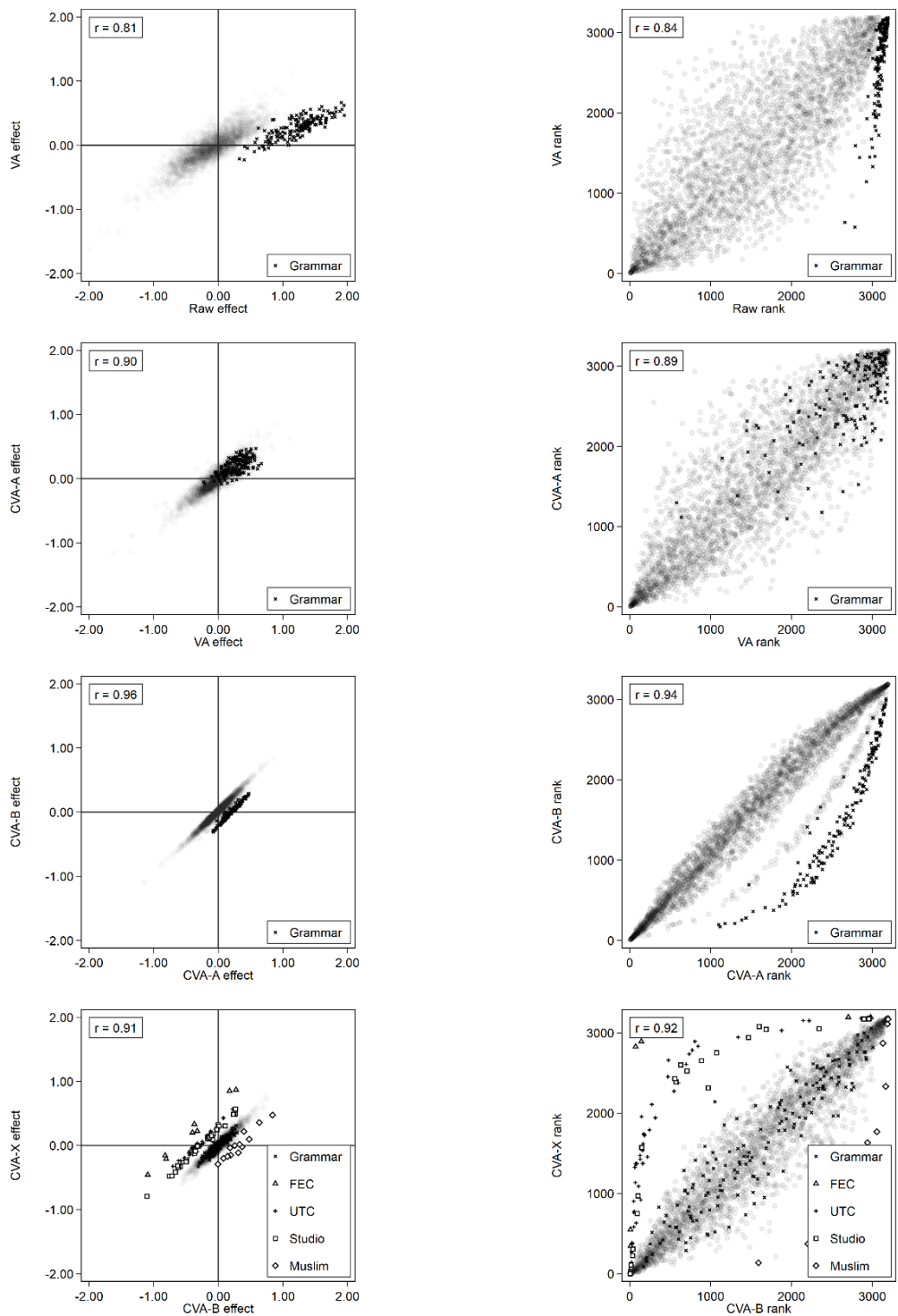

Notes.

Sample size = 3,196 schools. VA = value-added; CVA = contextual value-added.

**Figure 3.**

Line plots of the percentage of schools classified as having "moderate" or "large" school effects plotted against the five original school value-added models and plotted separately by the nature of the adjustment for student prior achievement in these models.

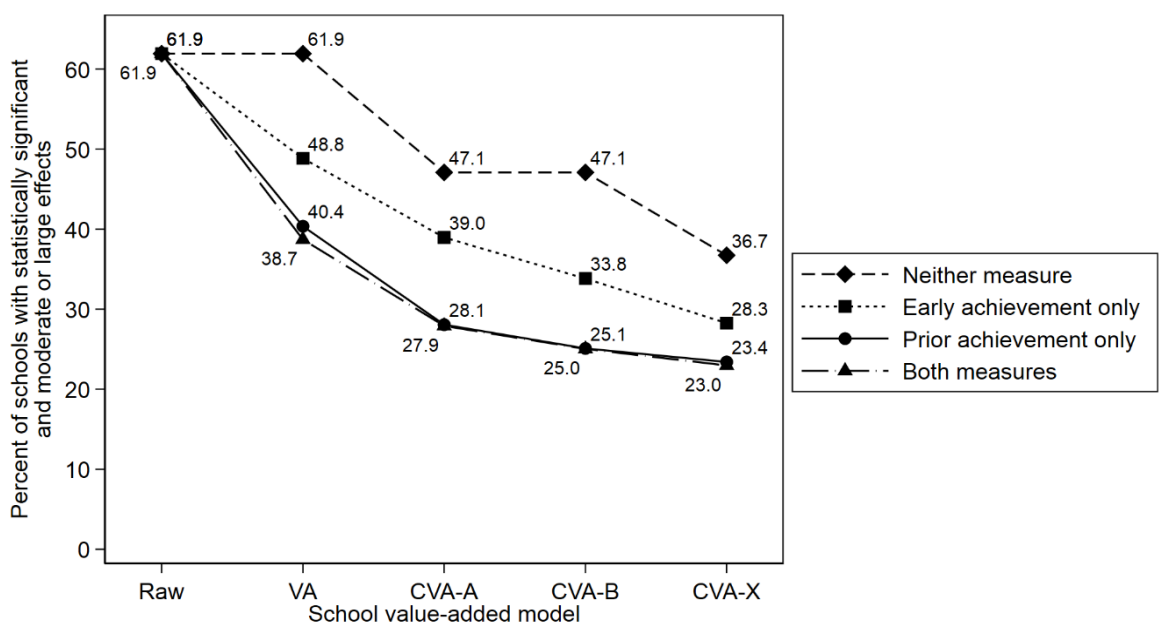

Notes.

Sample size = 3,196 schools. VA = value-added; CVA = contextual value-added. "Moderate" or "large" effects are defined as statistically significant effects greater than or equal to 0.2 SD of student current achievement.

**Figure 4.**

Line plots of correlations between each set of school effects (varying by the nature of their adjustment for student prior and early achievement) and the five original school value-added models.

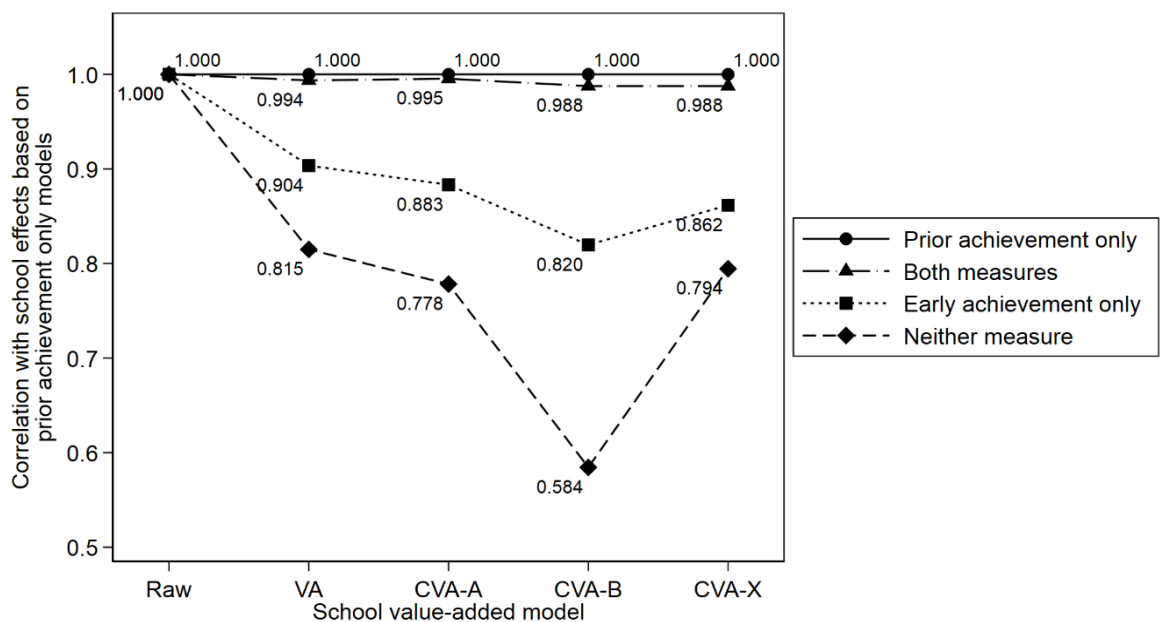

Notes

Sample size = 3,196 schools. VA = value-added; CVA = contextual value-added.

# About the authors

**George Leckie, Professor**

**University of Bristol**

George Leckie is a Professor of Social Statistics and Co-Director of the Centre for Multilevel Modelling at the School of Education, University of Bristol, UK. His methodological interests are in the development, application, and dissemination of multilevel and related models to analyse educational and other complex clustered and longitudinal data. His substantive interests focus on design, analysis, and communication issues surrounding school performance measures and league tables, especially the use of value-added models for estimating school effects on student achievement for accountability and choice purposes.

**Lucy Prior, Dr**

**University of Bristol**

Lucy Prior is a Research Associate in Quantitative Methods in Education at the School of Education, University of Bristol, UK. Her research interests include multilevel modelling, school performance measures, league tables, health geography and neighbourhood deprivation.

**Supplemental File**

In this Supplemental File, Tables S1-S7 report school effect variances and correlations for school value-added models reported in seven previous comparison studies that we focus on in the current study (Leckie & Goldstein, 2019; Marks, 2017; Marks, 2021; Muñoz-Chereau & Thomas, 2016; Thomas, 2001; Thomas & Mortimore, 1996; Timmermans et al., 2011). References can be found in the article reference list. Table S8 reproduces the school effect variances and correlations from the current study to facilitate comparisons with these earlier studies. Not all studies compare all models and so many of the tables have some empty cells. Furthermore, even for models that are compared, not all studies report the school variances or correlations in which case we present an "?" in the relevant table cells. Where studies present multiple sets of results corresponding to different phases of schooling, we pick the phase closest to the secondary phase focussed on in the current study. All school variances are derived from the estimated school variance model parameter. All school correlations are the sample correlations between the predicted school effects. As the student current achievement scales vary across previous studies, we translate all reported school variances to the values they would show if student current achievement had been standardized to have a SD of 1 prior to fitting the models.

**Table S1.**

School effect variances and correlations between school value-added models reported by Leckie and Goldstein (2019).

|        | Variance | Correlation |       |       |       |       |     |
|--------|----------|-------------|-------|-------|-------|-------|-----|
|        |          | Raw         | VA    | CVA-A | CVA-B | CVA-X | CA  |
| Raw    |          | 1           |       |       |       |       |     |
| VA     | 0.061    | 0.75        | 1     |       |       |       |     |
| CVA-A  | 0.047    | 0.61        | 0.91  | 1     |       |       |     |
| CVA-B  |          |             |       |       | 1     |       |     |
| CVA-X  |          |             |       |       |       | 1     |     |
| CA     |          |             |       |       |       |       | 1   |

Notes.

See Leckie and Goldstein (2019) for full details.

The education system is England.

The value-added phase is secondary schooling from Year 6 to Year 11.

Sample size is approximately 3000 schools.

VA = value-added; CVA = contextual value-added; CA = contextualised attainment.

The VA model is Progress 8.

The CVA-A model is Adjusted Progress 8.

Variances are reproduced from Table 1. We square the reported school SDs.

Correlations are reproduced from Figure 2. We reproduce the Pearson correlations presented in the top row.

**Table S2.**

School effect variances and correlations between school value-added models reported by Marks (2017).

|       | Variance | Correlation |       |       |       |       |     |
|-------|----------|-------------|-------|-------|-------|-------|-----|
|       |          | Raw         | VA    | CVA-A | CVA-B | CVA-X | CA  |
| Raw   |          | 1           |       |       |       |       |     |
| VA    | 0.013    |             | 1     |       |       |       |     |
| CVA-A |          |             |       | 1     |       |       |     |
| CVA-B |          |             |       |       | 1     |       |     |
| CVA-X |          |             |       |       |       | 1     |     |
| CA    | 0.057    |             | 0.85  |       |       |       | 1   |

Notes.

See Marks (2017) for full details. The notes below refer to that study.

The education system is Australia.

We focus on the results presented for secondary schooling value-added phase from Year 5 to Year 9.

We average results across the five current achievement subjects.

Sample size is approximately 70 schools.

VA = value-added; CVA = contextual value-added; CA = contextualised attainment.

The VA model is Model 3.

The CVA-A model is Model 1.

Variances are reproduced from Table 3. We average across the five subjects.

Correlations are reproduced from Table 5. We average across the five subjects.

**Table S3.**

School effect variances and correlations between school value-added models reported by Marks (2021).

|        | Variance | Correlation |       |       |       |       |     |
|--------|----------|-------------|-------|-------|-------|-------|-----|
|        |          | Raw         | VA    | CVA-A | CVA-B | CVA-X | CA  |
| Raw    | 0.300    | 1           |       |       |       |       |     |
| VA     | 0.054    | 0.89        | 1     |       |       |       |     |
| CVA-A  | 0.013    | 0.46        | 0.72  | 1     |       |       |     |
| CVA-B  | 0.011    | 0.24        | 0.54  | 0.96  | 1     |       |     |
| CVA-X  |          |             |       |       |       | 1     |     |
| CA     |          |             |       |       |       |       | 1   |

Notes.

See Marks (2021) for full details. The notes below refer to that study.

The education system is Australia.

We focus on the results presented for secondary schooling value-added phase from Year 5 to Year 9.

We average results across the five current achievement subjects.

Sample size is approximately 3000 schools.

VA = value-added; CVA = contextual value-added; CA = contextualised attainment.

The Raw model is Model 0.

The VA model is Model 1.

The CVA-A model is Model 3.

The CVA-B model is Model 4

Variances are reproduced from Table 5. We average across the five subjects and rescale the result using the information on the five current achievement score variances given in Table 1.

Correlations are reproduced from Table 7. We average across the five subjects.

**Table S4.**

School effect variances and correlations between school value-added models reported by Muñoz-Chereau & Thomas (2016).

|         | Variance | Correlation |     |       |       |       |     |
|---------|----------|-------------|-----|-------|-------|-------|-----|
|         |          | Raw         | VA  | CVA-A | CVA-B | CVA-X | CA  |
| Raw     | 0.373    | 1           |     |       |       |       |     |
| VA      | 0.089    | ?           | 1   |       |       |       |     |
| CVA-A   |          |             |     | 1     |       |       |     |
| CVA-B   | 0.029    | ?           | ?   |       | 1     |       |     |
| CVA-X   |          |             |     |       |       | 1     |     |
| CA      | 0.330    | ?           | ?   |       | ?     |       | 1   |

Notes.

See Muñoz-Chereau & Thomas (2016) for full details. The notes below refer to that study.

The education system is Chile.

The value-added phase is secondary schooling from Year 8 to Year 10.

We focus on the results presented for Mathematics.

Sample size is approximately 2300 schools.

VA = value-added; CVA = contextual value-added; CA = contextualised attainment.

The Raw model is Raw.

The VA model is VA

The CVA-A model is CVA.

The CVA-B model is CA.

Variances are reproduced from Table 3. We rescale the presented school variances by dividing through by the implied overall variance for student current achievement.

Correlations were not reported so appear as "?" here.

**Table S5.**

School effect variances and correlations between school value-added models reported by Thomas (2001).

|       | Variance | Correlation |       |       |       |       |    |
|-------|----------|-----|----|-------|-------|-------|----|
|       |          | Raw | VA | CVA-A | CVA-B | CVA-X | CA |
| Raw   | ?        | 1   |    |       |       |       |    |
| VA    | ?        | ?   | 1  |       |       |       |    |
| CVA-A | ?        | ?   | ?  | 1     |       |       |    |
| CVA-B | ?        | ?   | ?  | ?     | 1     |       |    |
| CVA-X |          |     |    |       |       | 1     |    |
| CA    | ?        | ?   | ?  | ?     | ?     |       | 1  |

Notes.

See Thomas (2001) for full details. The notes below refer to that study.

The education system focused on is primarily England.

VA = value-added; CVA = contextual value-added; CA = contextualised attainment.

The Raw model is Model 1.

The VA model is Models 3, 4, 5.

The CVA-A model is Models 6.

The CVA-B model is Model 7, 8

The CA model is Model 2.

Variances and correlations were not reported in the article as the article discussed these models at a higher, summary level of analysis as they explored models fitted to multiple student current achievement measures across multiple education systems.

**Table S6.**

School effect variances and correlations between school value-added models reported by Thomas and Mortimore (1996).

|        | Variance | Correlation |       |       |       |       |     |
|--------|----------|-------------|-------|-------|-------|-------|-----|
|        |          | Raw         | VA    | CVA-A | CVA-B | CVA-X | CA  |
| Raw    | 0.139    | 1           |       |       |       |       |     |
| VA     | 0.041    |             | 1     |       |       |       |     |
| CVA-A  | 0.037    | ?           | 0.94  | 1     |       |       |     |
| CVA-B  |          |             |       |       | 1     |       |     |
| CVA-X  |          |             |       |       |       | 1     |     |
| CA     | 0.058    | ?           | 0.77  | 0.79  |       |       | 1   |

Notes.

See Thomas and Mortimore (1996) for full details. The notes below refer to that study.

The education system is England.

The value-added phase is secondary schooling from Year 6 to Year 11.

Sample size is approximately 80 schools.

VA = value-added; CVA = contextual value-added; CA = contextualised attainment.

The Raw model is the Intercept model.

The VA model is the Prior attainment only model.

The CVA-A model is the Basic model.

The CA model is the No prior attainment model.

Variances are reproduced from Table 4. We rescale the presented school variances by dividing through by the implied overall variance for student current achievement.

Correlations are reproduced from Table 5.

**Table S7.**

School effect variances and correlations between school value-added models reported by Timmermans (2011).

|       | Variance | Correlation |      |       |       |       |    |
|-------|----------|------|------|-------|-------|-------|----|
|       |          | Raw  | VA   | CVA-A | CVA-B | CVA-X | CA |
| Raw   | 0.070    | 1    |      |       |       |       |    |
| VA    | 0.059    | 0.95 | 1    |       |       |       |    |
| CVA-A | 0.041    | 0.76 | 0.82 | 1     |       |       |    |
| CVA-B | 0.041    | 0.65 | 0.71 | 0.90  | 1     |       |    |
| CVA-X | 0.027    | 0.63 | 0.68 | 0.86  | 0.96  | 1     |    |
| CA    |          |      |      |       |       |       | 1  |

Notes.

See Timmermans (2011) for full details. The notes below refer to that study.

The education system is the Netherlands.

We focus on the results presented for higher general secondary education (HAVO).

Sample size is approximately 60 schools.

VA = value-added; CVA = contextual value-added; CA = contextualised attainment.

The Raw model is the Type 0 model.

The VA model is the Type AA model.

The CVA-A model is the Type A model.

The CVA-B model is the Type B model.

The CVA-X model is the Type X model.

Variances are reproduced from Table 6. We rescale the presented school variances by dividing through by the implied overall variance for student current achievement.

Correlations are reproduced from Table 7.

**Table S8.**

School effect variances and correlations between school value-added models reported by the current study.

|       | Variance | Correlation |       |       |       |       |     |
|-------|----------|-------------|-------|-------|-------|-------|-----|
|       |          | Raw         | VA    | CVA-A | CVA-B | CVA-X | CA  |
| Raw   | 0.226    | 1           |       |       |       |       |     |
| VA    | 0.069    | 0.81        | 1     |       |       |       |     |
| CVA-A | 0.043    | 0.70        | 0.90  | 1     |       |       |     |
| CVA-B | 0.039    | 0.49        | 0.84  | 0.96  | 1     |       |     |
| CVA-X | 0.032    | 0.41        | 0.71  | 0.87  | 0.91  | 1     |     |
| CA    |          |             |       |       |       |       | 1   |

Notes.

See the current study for full details. The notes below refer to that study.

The education system is England.

Sample size is approximately 3000 schools.

VA = value-added; CVA = contextual value-added; CA = contextualised attainment.

The value-added phase is secondary schooling from Year 6 to Year 11.

Variances are taken from Table 2. Row titled "Variance of school effects".

Correlations are taken from Table 3. Pearson correlations in lower triangle of table.